\documentclass[aps,prb,superscriptaddress,showpacs,twocolumn,bibnotes]{revtex4}
\usepackage{graphicx}
\usepackage{amssymb}
\usepackage{amsmath}

\newcommand{\bk}{{\bf k}}
\newcommand{\bq}{{\bf q}}

\newcommand{\bK}{{\bf K}}
\newcommand{\bp}{{\bf p}}

\newcommand{\br}{{\bf r}}
\newcommand{\bR}{{\bf R}}
\newcommand{\bJ}{{\bf J}}
\newcommand{\bz}{{\bf 0}}

\newcommand{\bd}{{\bf d}}

\newcommand{\bdelta}{{\boldsymbol\delta}}

\newcommand{\cotg}{{\mathop{\rm{cotg}}\nolimits}}
\renewcommand{\Im}{{\mathop{\rm{Im}}\nolimits\,}}
\renewcommand{\Re}{{\mathop{\rm{Re}}\nolimits\,}}

\newcommand{\kB}{k_{\mathrm{B}}}

\newcommand{\Ret}{{\mathrm{R}}}

\begin{document}

\title{Strain effect on the optical conductivity of graphene}

\author{F. M. D. Pellegrino}
\affiliation{Dipartimento di Fisica e Astronomia, Universit\`a di Catania,\\
Via S. Sofia, 64, I-95123 Catania, Italy}
\affiliation{Scuola Superiore di Catania, Via S. Nullo, 5/i, I-95123 Catania,
Italy}
\affiliation{CNISM, UdR Catania, I-95123 Catania, Italy}
\affiliation{INFN, Sez. Catania, I-95123 Catania, Italy}
\author{G. G. N. Angilella}
\email[Corresponding author. E-mail: ]{giuseppe.angilella@ct.infn.it}
\affiliation{Dipartimento di Fisica e Astronomia, Universit\`a di Catania,\\
Via S. Sofia, 64, I-95123 Catania, Italy}
\affiliation{Scuola Superiore di Catania, Via S. Nullo, 5/i, I-95123 Catania,
Italy}
\affiliation{INFN, Sez. Catania, I-95123 Catania, Italy}
\affiliation{CNISM, UdR Catania, I-95123 Catania, Italy}
\author{R. Pucci}
\affiliation{Dipartimento di Fisica e Astronomia, Universit\`a di Catania,\\
Via S. Sofia, 64, I-95123 Catania, Italy}
\affiliation{CNISM, UdR Catania, I-95123 Catania, Italy}

\date{\today}

\begin{abstract}

Within the tight binding approximation, we study the dependence of the
electronic band structure and of the optical conductivity of a graphene single
layer on the modulus and direction of applied uniaxial strain. While the Dirac
cone approximation, albeit with a deformed cone, is robust for sufficiently
small strain, band dispersion linearity breaks down along a given direction,
corresponding to the development of anisotropic massive low-energy excitations.
We recover a linear behavior of the low-energy density of states, as long as the
cone approximation holds, while a band gap opens for sufficiently intense
strain, for almost all, generic strain directions. This may be interpreted in terms
of an electronic topological transition, corresponding to a change of topology
of the Fermi line, and to the merging of two inequivalent Dirac points as a
function of strain. We propose that these features may be observed in the
frequency dependence of the longitudinal optical conductivity in the visible
range, as a function of strain modulus and direction, as well as of field
orientation.
\medskip
\pacs{%
78.40.Ri, 
62.20.-x, 
81.05.Uw 	
}
\end{abstract} 

\maketitle

\section{Introduction}

Graphene is an atomic thick single layer of carbon atoms in the $sp^2$
hybridization state, which in normal conditions crystallizes according to a
honeycomb lattice. The quite recent realization of sufficiently large graphene
flakes in the laboratory \cite{Novoselov:04,Novoselov:05} has stimulated an
enormous outburst of both experimental and theoretical investigation, due to its
remarkable mechanical and electronic properties, that make graphene an ideal
candidate for applications in nanoelectronics (see
Ref.~\onlinecite{CastroNeto:08} for a recent, comprehensive review).

Because the honeycomb lattice is composed of two interpenetrating triangular
sublattices, graphene is characterized by two bands, linearly dispersing at the
so-called Dirac points. Indeed, this is suggestive of the possibility of
observing relativistic effects typical of quantum electrodynamics in such a
unique condensed matter system \cite{Zhang:05,Berger:06}. The presence of
massless low-energy excitations also endows the density of states with a linear
dependence on energy at the Fermi level, which makes graphene a zero-gap
semiconductor. This in turn determines most of the peculiar transport properties
of graphene, including a minimal, finite conductivity in the clean limit at zero
temperature \cite{CastroNeto:08}, and a nearly constant conductivity over a
large frequency interval \cite{Gusynin:06,Stauber:08a}.

Graphene is also notable for its remarkable mechanical properties. In general,
nanostructures based on $sp^2$ carbon, such as also nanotubes and fullerenes,
are characterized by exceptional tensile strengths, despite their reduced
dimensionality. In particular, recent \emph{ab initio} calculations
\cite{Liu:07} as well as experiments \cite{Kim:09} have demonstrated that
graphene single layers can reversibly sustain elastic deformations as large as
20\%. In this context, it has been shown that Raman spectroscopy can be used as
a sensitive tool to determine the strain as well as some strain-induced
modifications of the electronic and transport properties of graphene
\cite{Ni:08,Mohiuddin:09}.

This opened the question whether applied strain could induce substantial
modifications of the band structure of graphene, such as the opening of a gap at
the Fermi level, thereby triggering a quantum phase transition from a semimetal
to a semiconductor. While earlier \emph{ab initio} calculations were suggestive
of a gap opening for arbitrary strain modulus and direction \cite{Gui:08}, both
tight-binding models \cite{Pereira:08a} as well as more accurate \emph{ab
initio} calculations \cite{Ribeiro:09} point towards the conclusion that the
strain-induced opening of a band gap in fact depends critically on the direction
of strain. On one hand, no gap opens when strain is applied in the armchair
direction, whereas on the other hand a sizeable strain modulus is required in
order to obtain a nonzero band~gap, for strain applied along a generic
direction. This result is in some sense consistent with the overall conviction
that the electron quantum liquid state in graphene, characterized by low-energy
massless excitations, lies indeed within some sort of `quantum protectorate',
\emph{i.e.} it is stable against sufficiently small, non-accidental
perturbations \cite{Laughlin:00}.

In this paper, we will be concerned on the effects induced by applied strain on
the optical conductivity of graphene. Among the various peculiar transport
properties of graphene, the conductivity as a function of frequency $\omega$ and
wavevector $\bk$ in graphene has received considerable attention in the past
(see Ref.~\onlinecite{Gusynin:07} for a review). The optical conductivity has
been derived within the Dirac-cone approximation \cite{Gusynin:06}, and within a
more accurate tight-binding approximation also for frequencies in the visible
range \cite{Stauber:08a}. The effect of disorder has been considered by Peres
\emph{et al.} \cite{Peres:06}, and that of finite temperature by Falkovsky and
Varlamov \cite{Falkovsky:07a}. These studies are consistent with the
experimentally observed of a nearly constant conductivity of $\pi e^2 /2h$ over
a relatively broad frequency range \cite{Mak:08,Nair:08}. Such a result
demonstrates that impurities and phonon effects can be neglected in the visible
range of frequencies \cite{Mak:08}.

Although uniaxial strain will be included in a standard, non-interacting model
Hamiltonian at the tight-binding level, \emph{i.e.} through the introduction of
strain-dependent hopping parameters \cite{Pereira:08a}, this will nonetheless
capture the essential consequences of applied strain on the band structure of
graphene. In particular, the strain-induced modification of the band structure
at a fixed chemical potential may result in an electronic topological transition
(ETT) \cite{Lifshitz:60} (see Refs.~\onlinecite{Blanter:94,Varlamov:99} for
comprehensive reviews). In metallic systems, this corresponds to a change of
topology of the Fermi line with respect to an external parameter, such as the
concentration of impurities or pressure, and is signalled by the appearance of
singularites in the density of states and other derived thermodynamic and
transport properties. The effect of the proximity to an ETT is usually enhanced
in systems with reduced dimensionality, as is the case of graphene. Here, one of
the main consequences of applied of strain is that of moving the Dirac points,
\emph{i.e.} the points where the band dispersion relations vanish linearly, away
from the points of highest symmetry in the first Brillouin zone, and of
deforming their low-energy conical approximation. Moreover, for a sufficiently
large strain modulus and a for a generic strain direction, two inequivalent
Dirac points may merge, thus resulting in the opening of a band gap (at strains
larger than the critical one) along one specific direction across the degenerate
Dirac point. This results in a sublinear density of states exactly at the
transition, which in turn gives rise to an unusual magnetic field dependence of
the Landau levels \cite{Montambaux:08}. This may also be described as a quantum
phase transition, from a semimetal to a semiconductor state, of purely
topological origin \cite{Wen:07}, characterized by low-energy massless
quasiparticles developing a finite mass only along a given direction.

It may be of interest to note that similar effects have been predicted also for
other low-dimensional systems, and that their overall features are generic with
respect to their detailed crystal structure. In particular, a similar discussion
applies to some quasi-two-dimensional Bechgaard salts
\cite{Angilella:02,Goerbig:08}, as well as to cold atoms in two-dimensional
optical lattices \cite{Montambaux:08,Wunsch:08}, which have been proposed to
simulate the behavior of Dirac fermions \cite{Zhu:07}.

The paper is organized as follows. In Sec.~\ref{sec:model} we review the tight
binding model for strained graphene, discuss the location of the Dirac points,
and the occurrence of the ETTs, as a function of strain. In Sec.~\ref{eq:dos} we
derive the Dirac cone approximation for the band dispersions in the presence of
strain, and discuss the low-energy energy dependence of the density of states.
This is then generalized over the whole bandwidth beyond the cone approximation.
The formation of band gaps is discussed with respect to strain modulus and
direction. In Sec.~\ref{sec:conductivity} we present the main results of this
paper, concerning the optical conductivity of strained graphene, and relate the
occurrence of several singularities in the frequency dependence thereof to the
various ETTs. We summarize our conclusions and give directions for future
studies in Sec.~\ref{sec:conclusions}.

\section{Model}
\label{sec:model}

Within the tight-binding approximation, the Hamiltonian for the graphene
honeycomb lattice can be written as
\begin{equation}
H = \sum_{\bR,\ell} t_\ell a^\dag (\bR) b(\bR+\bdelta_\ell) + \mathrm{H.c.},
\label{eq:H}
\end{equation}
where $a^\dag (\bR)$ is a creation operator on the position $\bR$ of the A
sublattice, $b(\bR+\bdelta_\ell)$ is a destruction operator on a nearest
neighbor (NN) site $\bR+\bdelta_\ell$, belonging to the B sublattice, and 
$\bdelta_\ell$ are the vectors connecting a given site to its nearest neighbors,
their relaxed (unstrained) components being $\bdelta_1^{(0)} =
a(1,\sqrt{3})/2$,  $\bdelta_2^{(0)} = a(1,-\sqrt{3})/2$,  $\bdelta_3^{(0)} =
a(-1,0)$, with $a=1.42$~\AA, the equilibrium C--C distance in a graphene sheet
\cite{CastroNeto:08}. In Eq.~(\ref{eq:H}), $t_\ell \equiv t(\bdelta_\ell )$,
$\ell=1,2,3$, is the hopping parameter between two NN sites. In the absence of
strain they reduce to a single constant, $t_\ell \equiv t_0$, with $t_0 =
-2.8$~eV (Ref.~\onlinecite{Reich:02}). 

\begin{figure}[t]
\centering
\includegraphics[height=0.9\columnwidth,angle=-90]{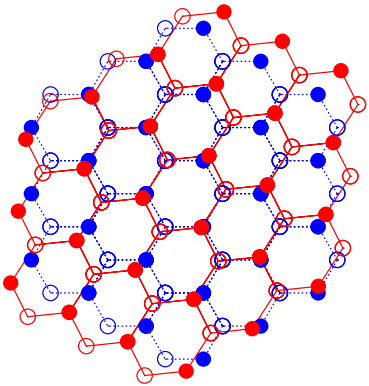}
\caption{(Color online) Schematic representation of the strained (red)
\emph{vs} unstrained (blue) honeycomb lattice, for $\theta=\pi/4$. Open (closed)
symbols refer to the A (B) sublattices, respectively.}
\label{fig:strain}
\end{figure}

In terms of the strain tensor \cite{Pereira:08a}
\begin{equation}
{\boldsymbol\varepsilon} = \varepsilon
\begin{pmatrix}
\cos^2 \theta -\nu \sin^2 \theta & (1+\nu)\cos\theta\sin\theta \\
(1+\nu)\cos\theta\sin\theta & \sin^2 \theta -\nu \cos^2 \theta
\end{pmatrix} ,
\label{eq:strainmat}
\end{equation}
the deformed lattice distances are related to the relaxed ones by
\begin{equation}
\bdelta_\ell = (\mathbb{I} +  {\boldsymbol\varepsilon}) \cdot \bdelta^{(0)}_\ell
.
\label{eq:straintransform}
\end{equation}
In Eq.~(\ref{eq:strainmat}), $\theta$ denotes the angle along which the strain
is applied, with respect to the $x$ axis in the lattice coordinate system,
$\varepsilon$ is the strain modulus, and $\nu=0.14$ is Poisson's ratio, as
determined from \emph{ab initio} calculations for graphene \cite{Farjam:09}, to
be compared with the known experimental value $\nu=0.165$ for graphite
\cite{Blakslee:70}. The special values $\theta=0$ and $\theta=\pi/6$ refer to
strain along the armchair and zig~zag directions, respectively.
Fig.~\ref{fig:strain} shows a schematic representation of the strained graphene
sheet, along the generic direction $\theta=\pi/4$, for definiteness.

Eqs.~(\ref{eq:strainmat}) and (\ref{eq:straintransform}) rely on the 
assumption that the lattice structure of graphene responds elastically
to applied strain, so that the effect of strain on the electronic properties
can be studied straightforwardly. The robustness of the elastic picture
is confirmed by recent atomistic simulations which have been compared with
available experiments \cite{Cadelano:09}.
It should be however mentioned that further detailed
calculations \cite{Chung:06,Liu:07,Ertekin:09} show that the formation
of topological defects under strain, such as Stone-Wales defects, 
can even induce structural phase transitions both in graphene and nanotubes.

Let ${\bf a}_\ell$ (${\bf a}^{(0)}_\ell$) the strained (unstrained) basis
vectors of the direct lattice, and ${\bf b}_\ell$ (${\bf b}^{(0)}_\ell$) the
strained (unstrained) basis vectors of the reciprocal lattice, respectively,
with \cite{CastroNeto:08} ${\bf b}^{(0)}_\ell = (2\pi/3a)(1,\pm\sqrt{3})$. One
has ${\bf b}_\ell = \mathbb{T} \cdot {\bf b}^{(0)}_\ell$. Then, from
Eq.~(\ref{eq:straintransform}), it is straightforward to show that $\mathbb{T} =
(\mathbb{I}+\boldsymbol{\varepsilon})^{-1}$. Since the wavevectors with and
without applied strain are connected by such a bijective tranformation, we can
safely work in the unstrained Brillouin zone (1BZ), \emph{i.e.}
$\bk\equiv\bk^{(0)}$.

It is useful to introduce the (complex) structure factor in momentum space, as
well as the NN hopping and overlap functions, respectively defined as
\begin{subequations}
\label{eq:gfg}
\begin{eqnarray}
\gamma_\bk &=& \sum_{\ell=1}^3 e^{i\bk\cdot\bdelta_\ell} ,\\
f_\bk &=& \sum_{\ell=1}^3 t_\ell e^{i\bk\cdot\bdelta_\ell} ,\\
g_\bk &=& \sum_{\ell=1}^3 s_\ell e^{i\bk\cdot\bdelta_\ell} .
\end{eqnarray}
\end{subequations}
Here, the strain-dependent overlap parameters $s_\ell$ are a generalization of
the band asymmetry parameter $s_0$ of Ref.~\onlinecite{Pellegrino:09}, and are
defined as
\begin{equation}
s_\ell = \int d\br \phi(\br) \phi(\br+\bdelta_\ell) 
= \exp \left( - \frac{\delta_\ell^2}{4\sigma_g^2} \right).
\end{equation}
Here, $\phi(\br) = (1/\sqrt{\pi}\sigma_g ) \exp(- r^2 /2\sigma_g^2 )$ is a
normalized gaussian pseudoatomic wavefunction, with $1/\sigma_g =
Z_g/2\sqrt{3}a$ (Refs.~\onlinecite{Bena:09,Pellegrino:09}), and the value $Z_g =
11.2$ is fixed by the condition that the relaxed overlap parameter be $s_0 =
0.07$ (Refs.~\onlinecite{Reich:02,Pellegrino:09}). Correspondingly, the hopping
parameters $t_\ell$ are defined as the transition amplitudes of the
single-particle Hamiltonian, $H_1 = -(\hbar^2/2m)\nabla^2 - Ze^2/r$, between two
lattice sites being $\bdelta_\ell$ apart from each other. Here, $Z$ is chosen so
that $t_\ell = t_0$ in the unstrained limit. One finds
\begin{equation}
t_\ell = \left[ 
\frac{\hbar^2}{2m\sigma_g^2} \left( 1 + \frac{\delta_\ell^2}{4\sigma_g^2} \right) -
\frac{Ze^2\sqrt{\pi}}{\sigma_g} I_0 \left( \frac{\delta_\ell^2}{8\sigma_g^2} \right)
\right]  s_\ell,
\end{equation}
where $I_0 (x)$ is a modified Bessel function of the first kind \cite{GR}. One
finds $dt_\ell /d\delta_\ell = 7.6$~eV/\AA{} for $\varepsilon=0$, which is
comparable with the value $6.4$~eV/\AA{} obtained in
Ref.~\onlinecite{Pereira:08a} within Harrison's approach\cite{Harrison:80}. In
the unstrained limit ($\varepsilon=0$), Eqs.~(\ref{eq:gfg}) reduce to $f_\bk \to
t_0 \gamma_\bk$ and $g_\bk \to s_0 \gamma_\bk$, respectively.

Within the tight binding approximation, the energy dispersion relations can be
obtained as the solutions $E_{\bk\lambda}$ of the generalized eigenvalue
problem
\begin{equation}
H_\bk {\bf u}_{\bk\lambda} = E_{\bk\lambda} S_\bk {\bf u}_{\bk\lambda} ,
\label{eq:eigen}
\end{equation}
where
\begin{subequations}
\begin{eqnarray}
H_\bk &=& \begin{pmatrix} 0 & f_\bk \\ f^\ast_\bk & 0 \end{pmatrix} , \\
S_\bk &=& \begin{pmatrix} 1 & g_\bk \\ g^\ast_\bk & 1 \end{pmatrix} .
\end{eqnarray}
\end{subequations}
One finds
\begin{equation}
E_{\bk\lambda} = \frac{-F_\bk \mp \sqrt{F_\bk^2 + 4 G_\bk
|f_\bk|^2}}{2G_\bk} ,
\label{eq:Ek}
\end{equation}
where $\lambda=1$ (minus sign) refers to the valence band, and $\lambda=2$
(plus sign) refers to the conduction band, and
\begin{subequations}
\begin{eqnarray}
F_\bk &=& g_\bk f_\bk^\ast + g_\bk^\ast f_\bk ,\\
G_\bk &=& 1 - |g_\bk |^2 .
\end{eqnarray}
\end{subequations}
The eigenvectors ${\bf u}_{\bk\lambda}$ in Eq.~(\ref{eq:eigen}) 
can be presented as
\begin{equation}
u_{\bk\lambda} = \begin{pmatrix} \cos\phi_{\bk\lambda} \\ e^{-i\theta_\bk} \sin
\phi_{\bk\lambda} \end{pmatrix} ,
\end{equation}
where $e^{i\theta_\bk} = f_\bk / |f_\bk|$, and
\begin{subequations}
\begin{eqnarray}
\cos\phi_{\bk\lambda} &=& - \frac{E_{\bk\bar{\lambda}}\sqrt{G_\bk}}
{\sqrt{|f_\bk|^2 + G_\bk E_{\bk\bar{\lambda}}^2}} ,\\
\sin\phi_{\bk\lambda} &=& - \frac{|f_\bk|}{\sqrt{|f_\bk|^2 +
G_\bk E_{\bk\bar{\lambda}}^2}} ,
\end{eqnarray}
\end{subequations}
with $\cos(\phi_{\bk,1}-\phi_{\bk,2})=0$. In the limit of no strain, one finds
$\phi_{\bk,1} \to 3\pi/4$ and $\phi_{\bk,2} \to \pi/4$.
Here and below, $\bar{\lambda}=2$ when $\lambda=1$, and 
vice~versa.

As already observed in Ref.~\onlinecite{Pellegrino:09}, a nonzero value of the
overlap parameters $s_\ell$ endows the conduction and valence bands with a
finite degree of asymmetry, which is here increasing with increasing modulus of
applied strain and, in general, anisotropic, depending on the direction of the
applied strain. In the unstrained limit ($\varepsilon=0$), one recovers the band
dispersions of Ref.~\onlinecite{Pellegrino:09}, $E_{\bk\lambda} \to \pm t_0
|\gamma_\bk| / (1\pm s_0 |\gamma_\bk|)$ (with $t_0 <0$).

The band dispersion relations $E_{\bk\lambda}$, Eq.~(\ref{eq:Ek}), are
characterized by Dirac points, \emph{i.e.} points in $\bk$-space around which
the dispersion is linear, when $f_\bk = 0$. As a function of strain, such a
condition is satisfied by two inequivalent points $\pm\bk_D$ only
when the `triangular inequalities'
\begin{equation}
|t_{\ell_1} -t_{\ell_2} | \leq |t_{\ell_3} | \leq |t_{\ell_1} + t_{\ell_2} |
\label{eq:triangular}
\end{equation}
are fulfilled \cite{Hasegawa:06}, with $(\ell_1,\ell_2,\ell_3)$ a permutation of
$(1,2,3)$. Around such points, the dispersion relations $E_{\bk\lambda}$ can be
approximated by cones, whose constant energy sections are ellipses.

The location of $\pm\bk_D$ in the reciprocal lattice satisfies
\begin{equation}
\cos\left( \bk_D \cdot (\bdelta_{\ell_1} - \bdelta_{\ell_2} )\right) =
\frac{t_{\ell_3}^2 - t_{\ell_1}^2 - t_{\ell_2}^2}{2t_{\ell_1}t_{\ell_2}} ,
\end{equation}
with $(\ell_1,\ell_2,\ell_3)$ a permutation of $(1,2,3)$. While in the
unstrained limit the Dirac points are located at the vertices of the 1BZ
(Ref.~\onlinecite{CastroNeto:08}), \emph{i.e.} $\bk_D \to \bK =
(2\pi/3a,2\pi/3\sqrt{3}a)$ and $-\bk_D \to \bK^\prime\equiv -\bK$, when either
of the limiting conditions in Eqs.~(\ref{eq:triangular}) is fulfilled as a
function of strain, say when $t_{\ell_3} = t_{\ell_1} + t_{\ell_2}$, the
would-be Dirac points coincide with the middle points of the sides of the 1BZ,
say $\bk_D \to M_{\ell_3}$. Here, $M_1 = \frac{\pi}{3a} (-1,-\sqrt{3})$,  $M_2 =
\frac{\pi}{3a} (-1,\sqrt{3})$, $M_3 = \frac{2\pi}{3a} (1,0)$. In this limit, the
dispersion relations cease to be linear in a specific direction, and the cone
approximation fails.

\begin{figure}[t]
\centering
\includegraphics[bb=55 162 540 655,clip,width=0.45\columnwidth]{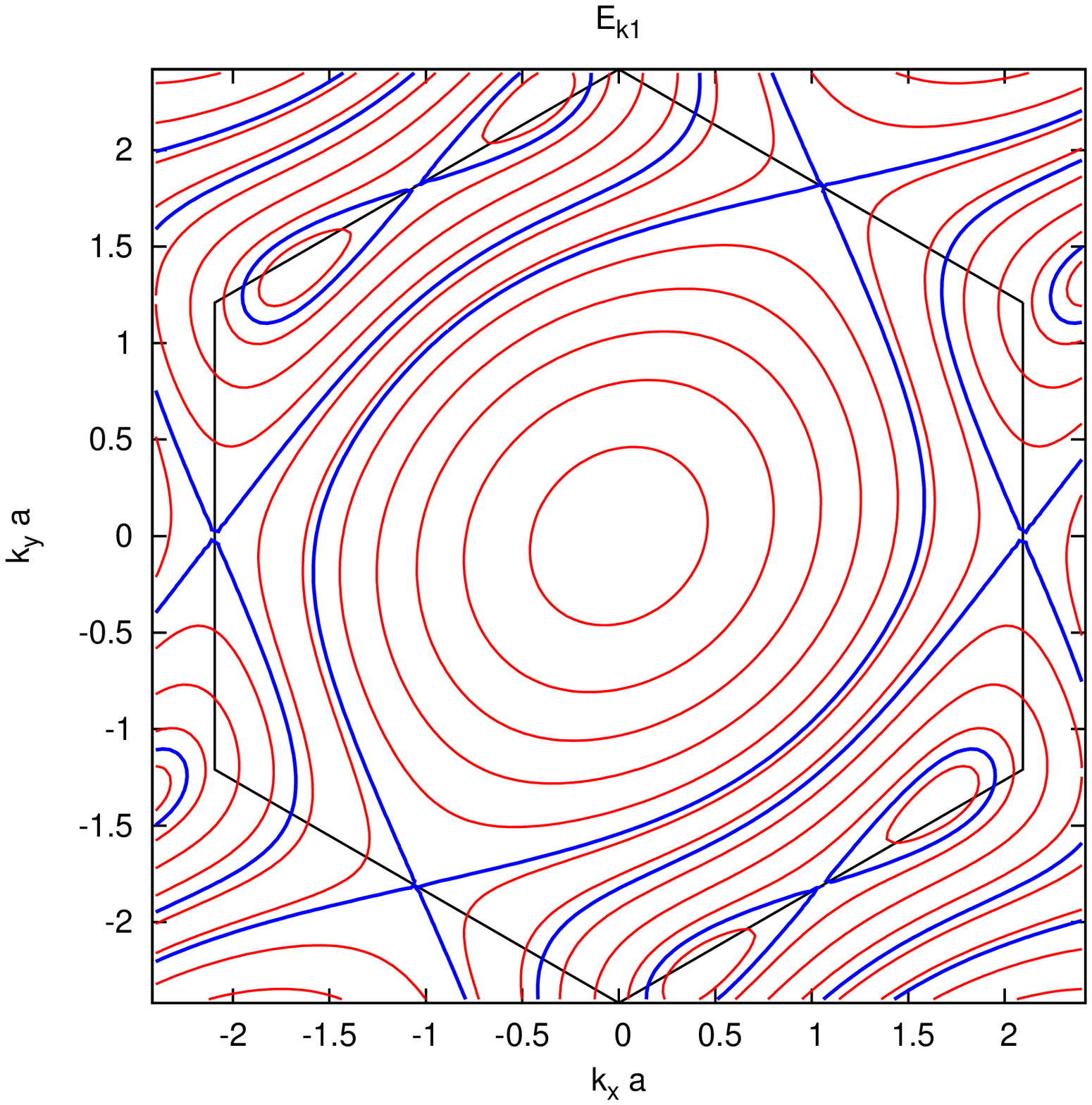}
\includegraphics[bb=55 162 540 655,clip,width=0.45\columnwidth]{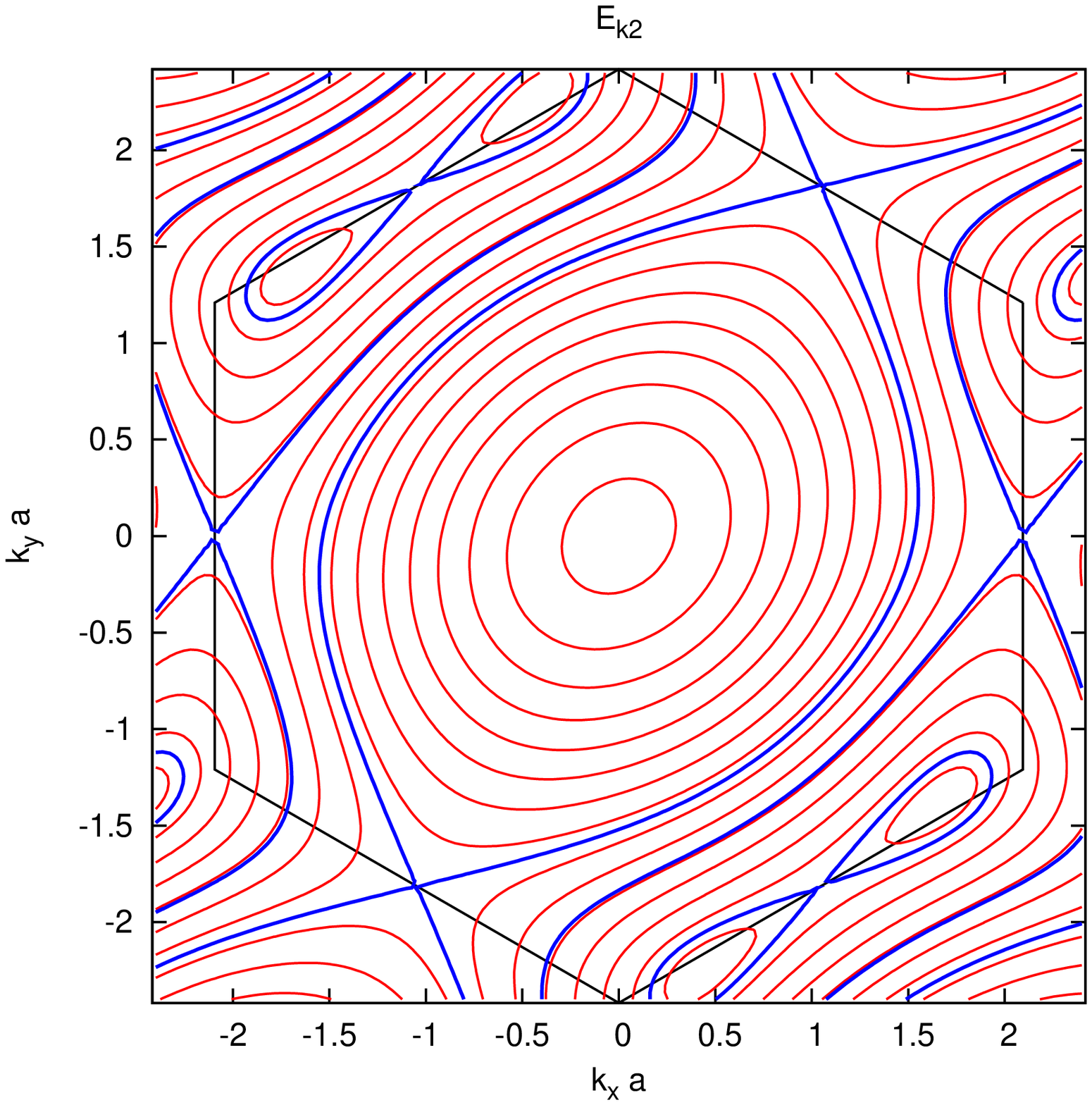}
\caption{(Color online) Contour plots of the dispersion relations within the 1BZ
for the valence band, $E_{\bk1}$ (left panel), and conduction band, $E_{\bk2}$
(right band), Eq.~(\ref{eq:Ek}). Here, we are depicting the situation
corresponding to a strain modulus of $\varepsilon=0.18$ along the generic
direction $\theta=\pi/4$. Solid blue lines are separatrix lines and occur at an
electronic topological transition, dividing groups of contours belonging to
different topologies. Either line passes through one of the critical points
$M_\ell$ ($\ell=1,2,3$), defined as the middle points of the 1BZ edge (solid
black hexagon).}
\label{fig:ETT}
\end{figure}

Fig.~\ref{fig:ETT} shows contour plots of $E_{\bk\lambda}$, Eq.~(\ref{eq:Ek}),
at constant energy levels. For fixed strain, each of these lines can be
interpreted as the Fermi line corresponding to a given chemical potential. One
may observe that the various possible Fermi lines can be grouped into four
families, according to their topology. In particular, from Fig.~\ref{fig:ETT}
one may distinguish among (1) closed Fermi lines around either Dirac point
$\pm\bk_D$ (and equivalent points in the 1BZ), (2) closed Fermi lines around
both Dirac points, (3) open Fermi lines, (4) closed Fermi lines around
$\Gamma=(0,0)$. The transition between two different topologies takes place when
the Fermi line touches the boundary of the 1BZ (solid black hexagon in
Fig.~\ref{fig:ETT}), and is marked by a separatrix line. It can be proved
explicitly that the Fermi line at the transition touches the boundary of the 1BZ
precisely at either of the hexagon sides midpoints $M_\ell$ ($\ell=1,2,3$),
defined above. This situation holds exactly also in the presence of overlap
($s_\ell \neq 0$), as can be proved within group theory \cite{Dresselhaus:08}.

Each separatrix line corresponds to an electronic topological transition (ETT)
\cite{Lifshitz:60,Blanter:94,Varlamov:99}, \emph{i.e.} a transition between two
different topologies of the Fermi line. An ETT can be induced by several
external parameters, such as chemical doping, or external pressure, or strain,
as in the present case. The hallmark of an ETT is provided by a kink in the
energy dependence of the density of states (DOS) of three-dimensional (3D)
systems, or by a logarithmic cusp (Van~Hove singularity) in the DOS of
two-dimensional (2D) systems. Besides being thoroughly studied in metals
\cite{Blanter:94,Varlamov:99}, the proximity to an ETT in quasi-2D cuprate
superconductors has been recently proposed to justify the nonmonotonic
dependence of the superconducting critical temperature on hole doping and other
material-dependent parameters, such as the next-nearest neighbor to nearest
neighbor hopping ratio \cite{Angilella:01}, as well as several normal state
properties, such as the fluctuation-induced excess Hall conductivity
\cite{Angilella:03g}. In particular, the role of epitaxial strain in inducing an
ETT in the cuprates has been emphasized \cite{Angilella:02d}. Due to the overall
$C_{2v}$ symmetry of the underlying 2D lattice in the CuO$_2$ layers, at most
two (usually degenerate) ETTs can be observed in the cuprates. Here, in the case
of strained graphene, characterized instead by $D_{3h}$ symmetry, we surmise the
existence of at most three, possibly degenerate, ETTs, whose effect on
observable quantities may be evidenced by the application of sufficiently
intense strain along specific directions.

\section{Density of states}
\label{eq:dos}

Under the conditions given by Eqs.~(\ref{eq:triangular}), the band dispersions,
Eqs.~(\ref{eq:Ek}), can be expanded as $E_{\bq\lambda}\equiv E_{\bk\lambda}$
around either Dirac point, say $\bk=\bk_D+\bq$, as:
\begin{equation}
E_{\bq\lambda} =
\frac{-\bq\cdot\bd\mp\sqrt{(\bq\cdot\bd)^2 + 4G_{\bk_D} |\bq\cdot\nabla
f_{\bk_D} |^2}}{2G_{\bk_D}} ,
\label{eq:cone}
\end{equation}
where
\begin{equation}
\bd = g_{\bk_D} \nabla f_{\bk_D}^\ast + g_{\bk_D}^\ast \nabla f_{\bk_D} .
\label{eq:hK}
\end{equation}
Eq.~(\ref{eq:cone}) defines a cone, whose section $E_{\bq\lambda} = E$ at a
constant energy level $E$ is an ellipse. Its equation can be cast in canonical
form as
\begin{equation}
\frac{(q_x -q_{x0})^2}{A^2} + \frac{(q_y -q_{y0})^2}{B^2} = E^2 ,
\label{eq:canonical}
\end{equation}
where the various parameters entering Eq.~(\ref{eq:canonical}) are defined in
App.~\ref{app:ellipse}. Making use of Eq.~(\ref{eq:canonical}), one can derive
the low-energy expansion of the density of states (DOS), which turns out to be linear
in energy,
\begin{equation}
\rho(E) = \rho_1 |E| ,
\label{eq:doslin}
\end{equation}
with
\begin{equation}
\rho_1 = \frac{4}{\pi} [(t_1^2 + t_2^2 + t_3^2)^2 -2(t_1^4 + t_2^4 +
t_3^4)]^{-1/2} ,
\label{eq:predos}
\end{equation}
where the factor of four takes into account for the spin and valley
degeneracies. 

\begin{figure}[t]
\centering
\includegraphics[height=0.95\columnwidth,angle=-90]{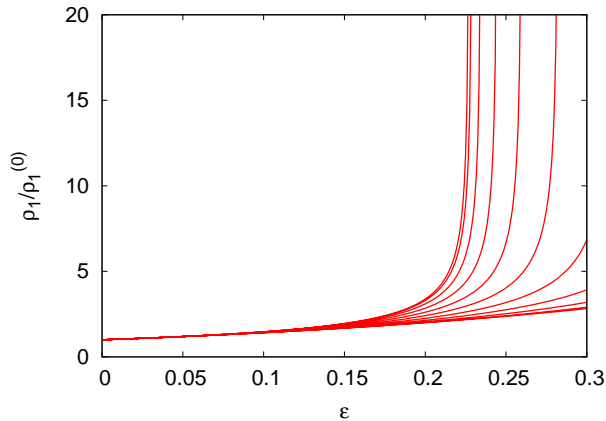}
\caption{(Color online) Showing the DOS prefactor $\rho_1$, Eq.~\ref{eq:predos},
normalized with respect to its value $\rho_1^{(0)}$ in the absence of strain, as
a function of the strain modulus $\varepsilon$, for various strain angles. The
strain direction $\theta$ increases from $\theta=0$ (armchair direction,
corresponding to the lowest curve) to $\theta=\pi/6$ (topmost curve). All other
cases can be reduced to one of these exploiting the symmetry properties of the
lattice.}
\label{fig:dosslope}
\end{figure}

Fig.~\ref{fig:dosslope} shows the prefactor $\rho_1$, Eq.~(\ref{eq:predos}), as
a function of the strain modulus $\varepsilon$, for various strain angles
$\theta$. One finds in general that $\rho_1$ increases monotonically with
increasing strain. Such a behavior suggests that applied strain may be used to
amplify the DOS close to the Fermi level. This, in particular, may serve as a
route to improve known methods to increase the carrier concentration in doped
graphene samples. When the equality sign in Eqs.~(\ref{eq:triangular}) is
reached, the prefactor $\rho_1$ in Eq.~(\ref{eq:predos}) diverges, meaning that
the cone approximation breaks down. In this case, the band dispersions still
vanish, but now quadratically along a specific direction through the would-be
Dirac point, and a nonzero gap in the DOS opens around $E=0$.

\begin{figure}[t]
\centering
\includegraphics[height=0.95\columnwidth,angle=-90]{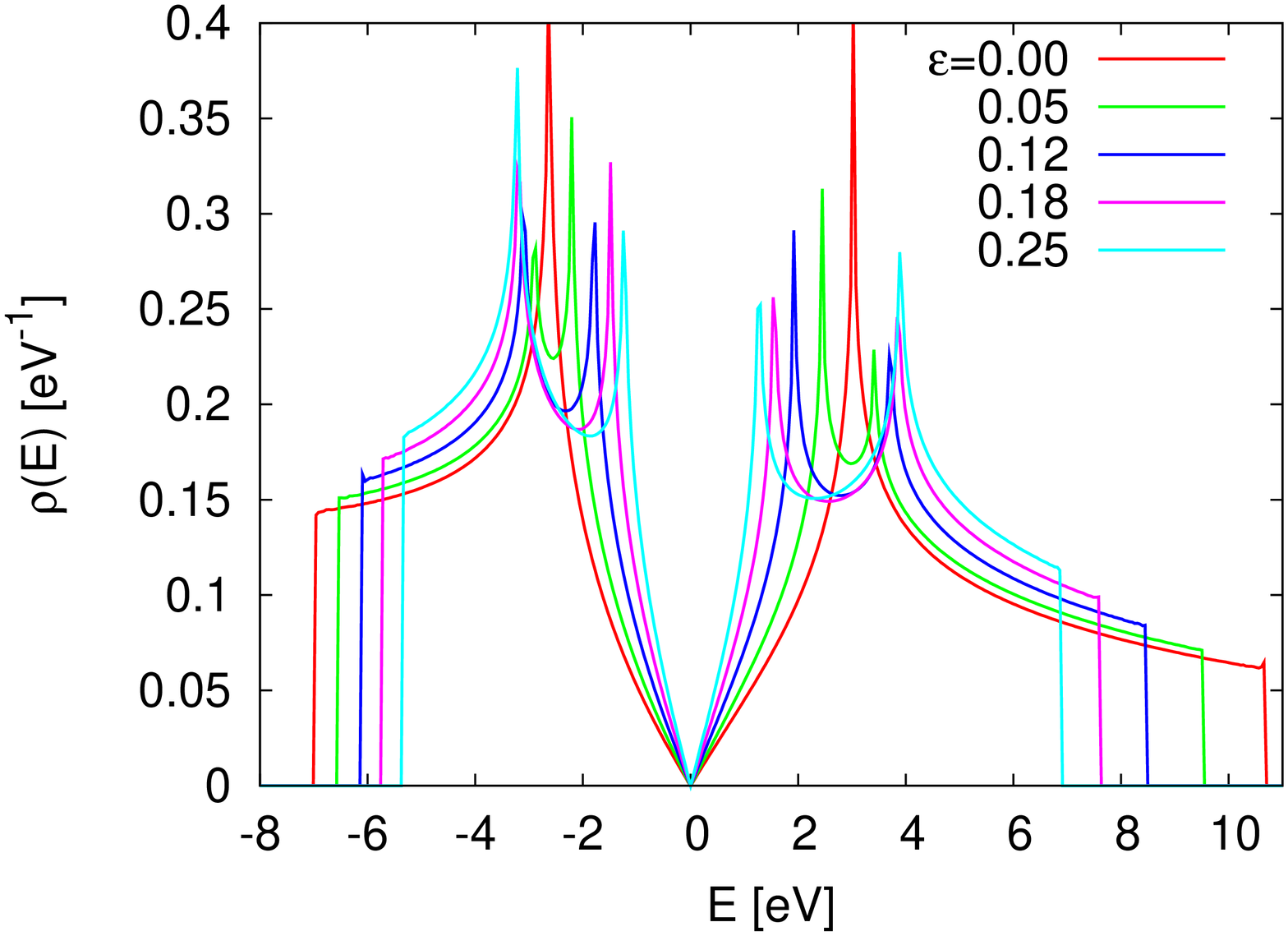}
\includegraphics[height=0.95\columnwidth,angle=-90]{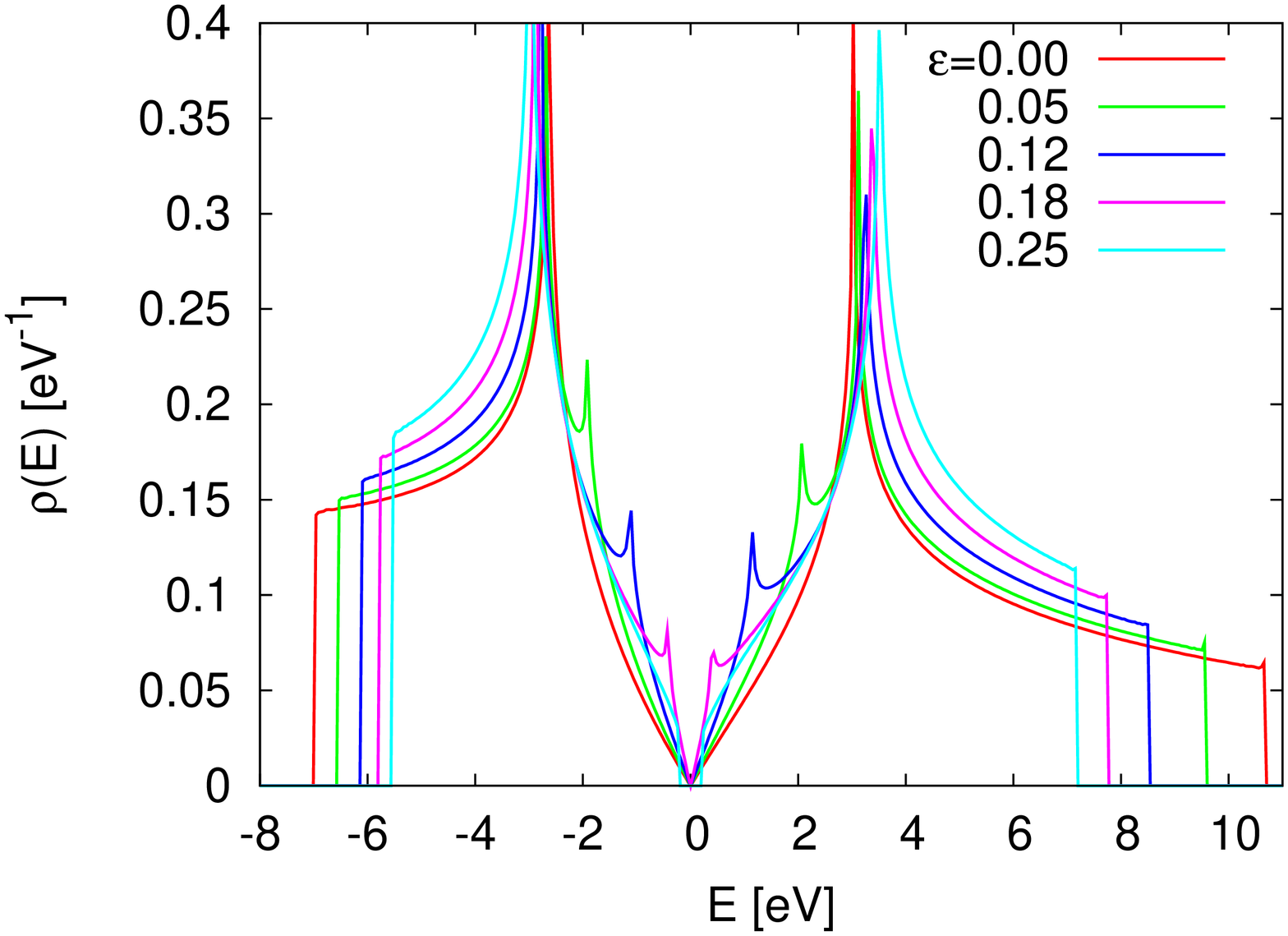}
\caption{(Color online) Energy dependence of the DOS over the whole bandwidth,
for increasing strain modulus $\varepsilon = 0-0.25$ and fixed strain direction
$\theta=0$ (top panel) and $\theta=\pi/6$ (bottom panel). In both cases, the DOS
slope close to the Fermi energy increases as a function of strain. However,
while the DOS remains gapless for $\theta=0$, a nonzero gap opens around $E=0$
at a critical strain for $\theta=\pi/6$.}
\label{fig:dosfulleps}
\end{figure}

\begin{figure}[t]
\centering
\includegraphics[height=0.95\columnwidth,angle=-90]{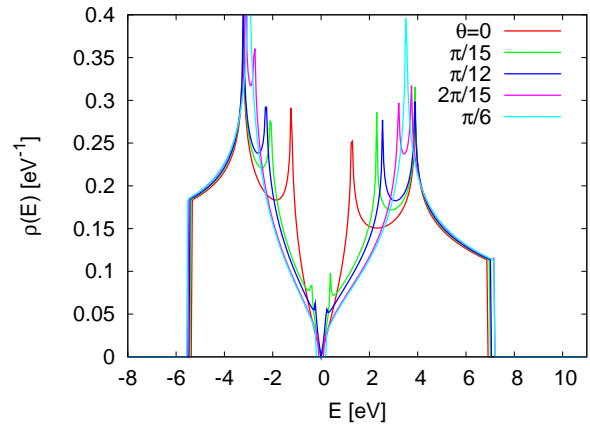}
\caption{(Color online) Energy dependence of the DOS over the full bandwidth,
for fixed strain modulus $\varepsilon=0.25$ and varying strain direction.}
\label{fig:dosfulltheta}
\end{figure}

This behavior is confirmed by the energy dependence of the DOS over the whole
bandwidth, as numerically evaluated from the detailed band dispersions,
Eq.~(\ref{eq:Ek}). In particular, Fig.~\ref{fig:dosfulleps} shows $\rho(E)$ for
increasing strain, at fixed strain angle $\theta=0$ (armchair) and
$\theta=\pi/6$ (zig~zag). In both cases, for sufficiently low values of the
strain modulus, the DOS depends linearly on $E$, according to
Eq.~(\ref{eq:doslin}), and the DOS slope increases with increasing strain, in
agreement with Eq.~(\ref{eq:predos}) and Fig.~\ref{fig:dosslope}. However, while
the spectrum remains gapless at all strains in the armchair case, a nonzero gap
is formed at a critical strain in the zig~zag case $\theta=\pi/6$, corresponding
to the breaking of the cone approximation at low energy. Such a behavior is
confirmed by Fig.~\ref{fig:dosfulltheta}, showing the dependence of the DOS over
the whole bandwidth, now at fixed strain modulus and varying strain angle.

At sufficiently high energies, beyond the linear regime, the DOS exhibits
Van~Hove singularities both in the valence and in the conduction bands. As
anticipated, these correspond to the occurrence of an ETT in the constant energy
contours of either band dispersion relation $E_{\bk\lambda}$, Eq.~(\ref{eq:Ek}).
As shown by Fig.~\ref{fig:dosfulleps}, the DOS is characterized by a single
logarithmic cusp in each band in the unstrained limit ($\varepsilon=0$), that is
readily resolved into two logarithmic spikes, both in the $\theta=0$ (armchair)
and in the $\theta=\pi/6$ (zig~zag) cases, as soon as the strain modulus
becomes nonzero ($\varepsilon>0$). The low-energy spike disappears as soon as a
gap is formed, corresponding to the breaking of the cone behavior around the
Dirac point. Fig.~\ref{fig:dosfulltheta} shows that the situation is
indeed richer, in that the application of sufficiently intense strain along
generic (\emph{i.e.} non symmetry-privileged) directions allows the development
of three logarithmic singularities in the DOS for each band,
corresponding to the three inequivalent ETTs described in
Section~\ref{sec:model}. Again, the lowest energy Van~Hove singularity
disappears into the gap edge when the energy spectrum ceases to be linear around
the Dirac points. This takes place when the Dirac points $\pm\bk_D$ tend to
either edge midpoint $M_\ell$ of the 1BZ ($\ell=1,2,3$). In this case, the
energy gap $\Delta_\ell$ can be found explicitly and, in the simple case of no
overlap ($s_\ell =0$), can be written as
\begin{subequations}
\begin{eqnarray}
\Delta_1 &=& 2 \sqrt{t_1^2 + t_2^2 + t_3^2 -2 t_1 t_2 -2 t_1 t_3 + 2 t_2 t_3}
,\\
\Delta_2 &=& 2 \sqrt{t_1^2 + t_2^2 + t_3^2 -2 t_1 t_2 +2 t_1 t_3 - 2 t_2 t_3}
,\\
\Delta_3 &=& 2 \sqrt{t_1^2 + t_2^2 + t_3^2 +2 t_1 t_2 -2 t_1 t_3 - 2 t_2 t_3}
, 
\end{eqnarray}
\end{subequations}
for $\ell=1,2,3$, respectively.

\begin{figure}[b]
\centering
\includegraphics[height=0.95\columnwidth,angle=-90]{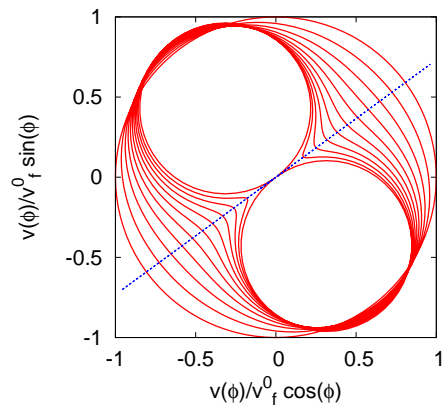}
\caption{(Color online) Polar plot of $v_\lambda (\phi)$ (with $\lambda=2$,
\emph{i.e.} for the conduction band) around $\bk_D$, $v_\lambda (\phi)$,
normalized with respect to its value in the absence of strain, $v^{(0)}_f$.
Strain is here applied at a generic fixed angle $\theta=\pi/4$. The anisotropy
of the Fermi velocity increases with increasing strain, until the shape of
$v_\lambda(\phi)$ breaks at $\varepsilon=0.28$. This corresponds to the
existence of a direction (solid blue line), Eq.~(\ref{eq:cotg}), along which the
dispersion relation $E_{\bq\lambda}$ displays a nonlinear character.}
\label{fig:vphi}
\end{figure}

Further insight into the anisotropical character of the low-energy cone dispersion
relations around the Dirac points, Eq.~(\ref{eq:cone}) can be obtained by
recasting them in polar coordinates $(q,\phi)$, where
$\bq=(q\cos\phi,q\sin\phi)$. One finds therefore $E_{\bq\lambda} = v_{\lambda}
(\phi) q$, the anisotropic prefactor $v_{\lambda} (\phi)$ depending on the Dirac
point around which one is actually performing the expansion. Fig.~\ref{fig:vphi}
shows $v_{\lambda} (\phi)$ for the conduction band ($\lambda=2$) centered around
$\bk_D$. One notices that applied strain increases the anisotropy of the $\phi$
dependence, until a critical value is reached, at which the cone approximation
breaks down. This corresponds to a nonlinear behavior of $E_{\bq\lambda}$ along
a specific direction $\phi_0$, characterized by the vanishing of $v_\lambda (\phi)$ and
given explicitly by
\begin{equation}
\cotg\phi_0 = -\frac{\sqrt{3}}{3} \frac{t_1 \mp t_2}{t_1 \pm t_2} ,
\label{eq:cotg}
\end{equation}
when $|t_3 | = |t_1 \mp t_2 |$ in Eqs.~(\ref{eq:triangular}), and to the opening
of a finite gap around zero energy in the DOS. In that case, the Fermi velocity
vanishes along a direction $\phi_0^\prime$ given by
\begin{equation}
\cotg \phi_0^\prime = \frac{(1+ \varepsilon_{11})\cotg\phi_0 -
\varepsilon_{12}}{(1+\varepsilon_{22}) - \varepsilon_{21} \cotg\phi_0} .
\end{equation}

\section{Optical conductivity}
\label{sec:conductivity}

The paramagnetic component of the density current vector in momentum space reads
\cite{Bruus:04,Pellegrino:09}
\begin{equation}
\tilde{\bJ}^\nabla (\bp^\prime) = -\frac{e}{2m} \int \frac{d\bp}{(2\pi)^2} (2\bp
+ \bp^\prime) c_\bp^\dag c_{\bp+\bp^\prime} ,
\end{equation}
where $c_\bp$ ($c^\dag_\bp$) are destruction (creation) operators in the plane
wave representation. In the homogeneous limit (zero transferred momentum,
$\bp^\prime=\bz$), one has \cite{Paul:03}
\begin{equation}
\tilde{\bJ}^\nabla (0) = \frac{e}{i\hbar} [H,\br]  = -e \dot \br,
\label{eq:Paul}
\end{equation}
where $H$ is the system's Hamiltonian.

Within linear response theory, the conductivity $\sigma$ is related to the
current-current correlation function through a Kubo formula
\begin{equation}
\sigma_{lm} (\mu,T;\omega) = \frac{ie^2 n}{m\omega}\delta_{lm} +
\frac{i}{\hbar\omega N A_{cell}} \tilde{\Pi}^\Ret_{lm} (0,0,\omega),
\end{equation}
where $n$ is the electron density, $T$ is the temperature, $\omega$ is the
frequency of the external electric field, $A_{cell}$ is the area of a primitive
cell, and $\tilde{\Pi}^\Ret_{lm} (\bp,\bp^\prime,\omega)$ is the $(l,m)$
component of the Fourier transform of the retarded current-current correlation
tensor. One is usually concerned with the dissipative part of the conductivity
tensor, \emph{i.e.} its real part. One has therefore
\begin{equation}
\sigma_{lm} (\mu,T;\omega) = - \frac{1}{\hbar\omega N A_{cell}} \Im
\tilde{\Pi}^\Ret_{lm} (0,0,\omega),
\end{equation}
where $\tilde{\Pi}^\Ret_{lm}$ is the retarded version of
\begin{equation}
\tilde{\Pi}_{lm} (\bp,\bp^\prime,\tau) = - \langle T_\tau [ \tilde{J}^\nabla_l
(\bp,\tau) \tilde{J}^\nabla_m (\bp^\prime,0) ] \rangle ,
\end{equation}
and $\tilde{J}^\nabla_l (\bp,\tau)$ denotes the Fourier transform of the
paramagnetic component of the current density vector, at the imaginary time
$\tau$. Projecting $\tilde{J}^\nabla_l (\bp,\tau)$ onto the tight-binding
states, and neglecting the Drude peak, one finds
\begin{widetext}
\begin{equation}
\sigma_{lm}(\omega) = \Re \frac{2i}{A_{cell} \hbar\omega} \frac{1}{N}
\sum_{\bk\lambda}
\left[
\left( \tilde{J}^\nabla_l (\bk) \right)_{\lambda\bar{\lambda}}
\left( \tilde{J}^\nabla_m (\bk) \right)_{\bar{\lambda}\lambda}
\frac{n_F (\xi_{\bk\bar{\lambda}}) - n_F (\xi_{\bk\lambda})}{\hbar\omega +
\xi_{\bk\lambda} - \xi_{\bk\bar{\lambda}} + i0^+} 
\right] ,
\end{equation}
where $\xi_{\bk\lambda} = E_{\bk\lambda}-\mu$ and $n_F (\xi)$ denotes the Fermi
function at temperature $T$. In the direction of the external field, \emph{i.e.}
for $l=m$, one finds
\begin{equation}
\frac{\sigma_{ll}(\omega)}{\sigma_0} =  \frac{2\pi}{\tau_0^2} 
\frac{\sinh(\frac{1}{2}\hbar\beta|\omega|)}{\hbar\omega}
\frac{1}{N} \sum_{\bk\lambda}
\left|  \tilde{j}^\nabla_l (\bk) \right|_{\lambda\bar{\lambda}}^2
F(\beta,\mu;\bk) \delta \left(\hbar\omega - (E_{\bk\lambda} -
E_{\bk\bar{\lambda}})\right) ,
\label{eq:sigma}
\end{equation}
\end{widetext}
where $\beta=(\kB T)^{-1}$ is the inverse temperature, $\tilde{J}^\nabla_l (\bk)
= e\frac{ta}{\hbar} \tilde{j}^\nabla_l (\bk)$, $\sigma_0 = \pi e^2/(2h)$ is
proportional to the quantum of conductivity, $\tau_0^{-2} = 16 t^2 / (3\sqrt{3}
\pi\hbar^2)$, and
\begin{subequations}
\begin{eqnarray}
F(\beta,\mu;\bk) &=& 2 e^{\beta(\bar{E}_\bk - \mu)} n_F (\xi_{\bk,1})
n_F(\xi_{\bk,2}) ,\\
\bar{E}_\bk &=& \frac{E_{\bk,1}+E_{\bk,2}}{2} ,
\end{eqnarray}
\end{subequations}
with $\bar{E}_\bk = 0$ when $g_\bk=0$ (no overlap).

\begin{figure}[t]
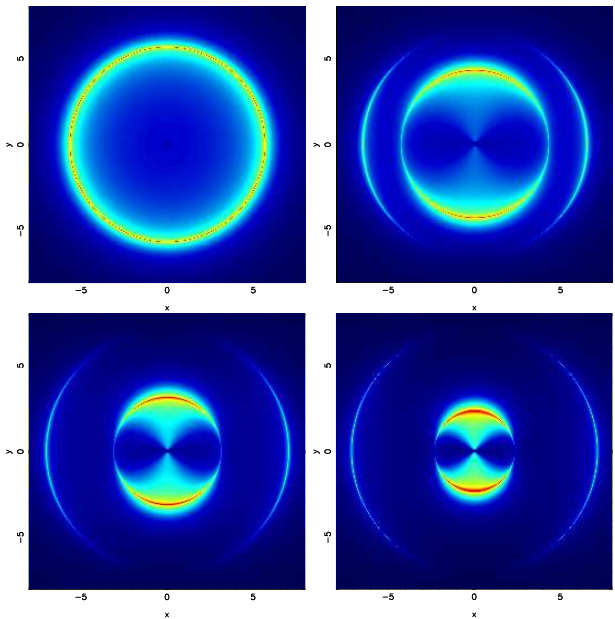

\begin{center}
\includegraphics[height=0.46\columnwidth,angle=-90]{sigmacontour0zoom_1.ps}
\includegraphics[height=0.46\columnwidth,angle=-90]{sigmacontour0zoom_2.ps}\\
\includegraphics[height=0.46\columnwidth,angle=-90]{sigmacontour0zoom_3.ps}
\includegraphics[height=0.46\columnwidth,angle=-90]{sigmacontour0zoom_4.ps}
\end{center}
\caption{(Color online) Polar plots of longitudinal optical conductivity
$\sigma_{ll}/\sigma_0$, Eq.~(\ref{eq:sigma}), as a function of frequency
$\omega>0$ (polar axis) and electric field orientation $\phi$ (azymuthal
direction). Here, we set $\mu=0$ and $\kB T=0.025$~eV. Strain is applied along
the $\theta=0$ (armchair) direction, and the strain modulus increases from left
to right, and from top to bottom ($\varepsilon=0,0.075,0.175,0.275$).}
\label{fig:sigmacontour0}
\end{figure}

\begin{figure}[t]
\centering
\includegraphics[height=0.9\columnwidth,angle=-90]{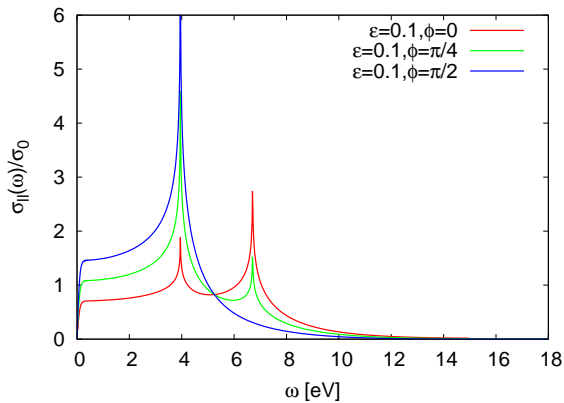}
\caption{(Color online) Longitudinal optical conductivity
$\sigma_{ll}/\sigma_0$, Eq.~(\ref{eq:sigma}), as a function of frequency
$\omega>0$, for fixed strain modulus $\varepsilon=0.1$ and strain direction
$\theta=0$ (armchair). Different lines refer to various orientations of the
electric field ($\phi=0,\pi/4,\pi/2$).}
\label{fig:sigma0}
\end{figure}

We have numerically evaluated the longitudinal optical conductivity $\sigma_{ll}
(\omega)$, Eq.~(\ref{eq:sigma}) as a function of frequency $\omega>0$ at fixed
temperature $\kB T = 0.025$~eV, for several strain moduli $\varepsilon$ and
directions $\theta$, as well as field orientations, here parametrized by the
angle $\phi$ between the applied electric field and the lattice $x$ direction.
Figs.~\ref{fig:sigmacontour0} and \ref{fig:sigma0} show our results in the case
of strain applied in the armchair direction ($\theta=0$).
Fig.~\ref{fig:sigmacontour0} shows a contour plot of the longitudinal optical
conductivity $\sigma_{ll}/\sigma_0$ as a function of frequency $\omega$ (radial
coordinate) and applied field angle (polar angle). In the relaxed limit
($\varepsilon=0$), $\sigma_{ll}/\sigma_0$ is isotropic with respect to the
applied field angle, and exhibits a maximum at a frequency that can be related to the
single Van~Hove singularity in the DOS (cf. Fig.~\ref{fig:dosfulleps}). Such a
maximum is immediately split into distinct maxima, in general, as soon as
the strain modulus $\varepsilon$ becomes nonzero. This can be interpreted in
terms of applied strain partly removing the degeneracy among the inequivalent
underlying ETTs. Such an effect is however dependent on the field direction
$\phi$, as is shown already by the anisotropic pattern developed by
$\sigma_{ll}/\sigma_0$ in Fig.~\ref{fig:sigmacontour0}, for $\varepsilon\neq0$.
Indeed, Fig.~\ref{fig:sigma0} shows plots of $\sigma_{ll}/\sigma_0$ as a
function of frequency for fixed strain modulus $\varepsilon=0.1$ and varying
field orientation $\phi=0-\pi/2$. The relative weight of the three maxima
depends on the relative orientation between strain and applied field.
Here and below, we consider the case $\mu=0$. A nonzero value of the
chemical potential would result in a vanishing conductivity below a cutoff at
$\omega\approx|\mu|$, smeared by finite temperature effects
\cite{Stauber:08a}.

\begin{figure}[t]
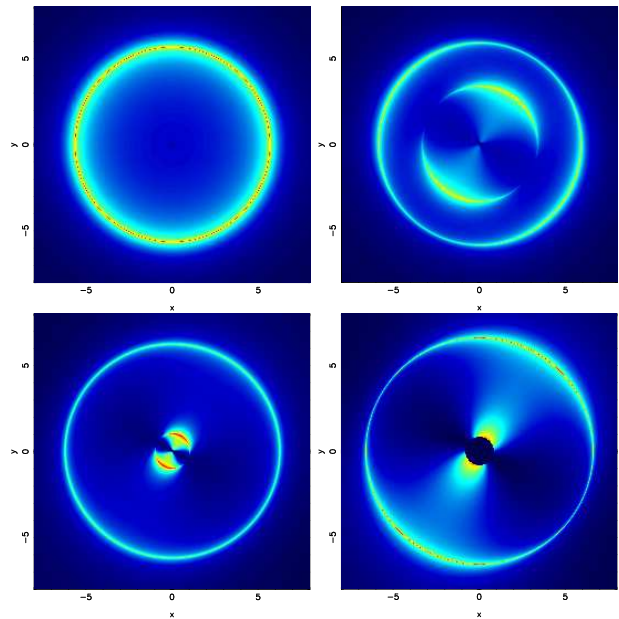

\begin{center}
\includegraphics[height=0.46\columnwidth,angle=-90]{sigmacontour30zoom_1.ps}
\includegraphics[height=0.46\columnwidth,angle=-90]{sigmacontour30zoom_2.ps}\\
\includegraphics[height=0.46\columnwidth,angle=-90]{sigmacontour30zoom_3.ps}
\includegraphics[height=0.46\columnwidth,angle=-90]{sigmacontour30zoom_4.ps}
\end{center}
\caption{(Color online) Same as Fig.~\ref{fig:sigmacontour0}, but for strain
applied the $\theta=\pi/6$ direction.}
\label{fig:sigmacontour30}
\end{figure}

\begin{figure}[t]
\centering
\includegraphics[height=0.9\columnwidth,angle=-90]{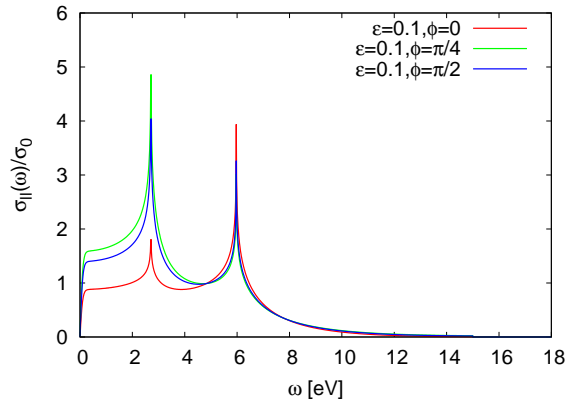}
\caption{(Color online) Same as Fig.~\ref{fig:sigma0}, but for strain applied the
$\theta=\pi/6$ direction.}
\label{fig:sigma30}
\end{figure}

An analogous behavior is recovered when strain is applied along the zig~zag
direction $\theta=\pi/6$, as shown in Figs.~\ref{fig:sigmacontour30} and
\ref{fig:sigma30}. Again, applied strain breaks down the original isotropy of
the optical conductivity with respect to the field orientation in the relaxed
case, with two maxima appearing as a function of frequency
(Fig.~\ref{fig:sigmacontour30}). The optical weight of the different maxima
depend in general by the relative orientation between strain and applied field.
While the presence of the two peaks can be traced back to the existence of
inequivalent ETTs, whose degeneracy is here removed by applied strain, the last
panel in Fig.~\ref{fig:sigmacontour30} shows that at a sufficiently large strain
modulus (here, $\varepsilon=0.275$), a gap opens in the low-energy sector of the
spectrum, which is signalled here by a vanishing optical conductivity (dark spot
at the origin in last panel of Fig.~\ref{fig:sigmacontour30}).

\begin{figure}[t]
\begin{center}
\includegraphics[height=0.46\columnwidth,angle=-90]{sigmacontour45zoom_1.ps}
\includegraphics[height=0.46\columnwidth,angle=-90]{sigmacontour45zoom_2.ps}\\
\includegraphics[height=0.46\columnwidth,angle=-90]{sigmacontour45zoom_3.ps}
\includegraphics[height=0.46\columnwidth,angle=-90]{sigmacontour45zoom_4.ps}
\end{center}
\caption{(Color online) Same as Fig.~\ref{fig:sigmacontour0}, but for strain
applied the $\theta=\pi/4$ direction.}
\label{fig:sigmacontour45}
\end{figure}

\begin{figure}[t]
\centering
\includegraphics[height=0.9\columnwidth,angle=-90]{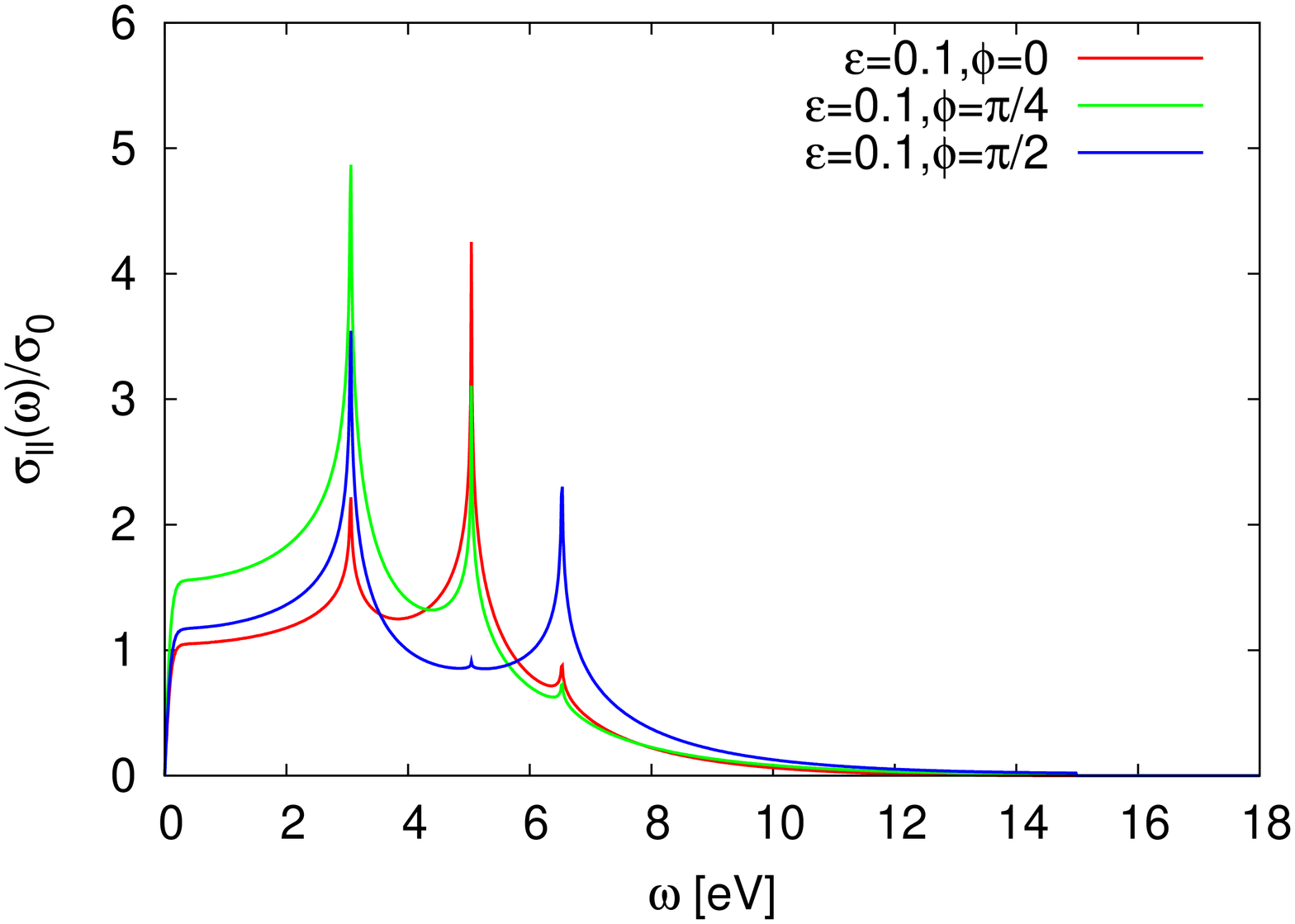}
\caption{(Color online) Same as Fig.~\ref{fig:sigma0}, but for strain applied the
$\theta=\pi/4$ direction.}
\label{fig:sigma45}
\end{figure}

Finally, Figs.~\ref{fig:sigmacontour45} and \ref{fig:sigma45} show the
longitudinal optical conductivity in the case of increasing strain applied along
a generic direction, \emph{viz.} $\theta=\pi/4$. Like in the previous cases,
applied strain removes the isotropy of $\sigma_{ll}/\sigma_0$ with respect to
the field orientation $\phi$. However, the degeneracy among the three
inequivalent ETTs is here lifted completely, and three peaks in general appear
in the longitudinal optical conductivity as a function of frequency, as shown
also by Fig.~\ref{fig:sigma45}. The redistribution of optical weight among the
three peaks is now more complicated, as it in general depends on both the strain
direction $\theta$ and the field orientation $\phi$.

\section{Conclusions}
\label{sec:conclusions}

We have discussed the strain dependence of the band structure, and derived the
strain and field dependence of the optical conductivity of graphene under
uniaxial strain. Within a tight-binding model, including strain-dependent
nearest neighbour hoppings and orbital overlaps, we have interpreted the
evolution of the band dispersion relations with strain modulus and direction in
terms of the proximity to several electronic topological transitions (ETT).
These correspond to the change of topology of the Fermi line as a function of
strain. In the case of graphene, one may distinguish among three distinct ETTs.
We also recover the evolution of the location of the Dirac points, which move
away from the two inequivalent symmetric points $\bK$ and $\bK^\prime$ as a
function of strain. For sufficiently small strain modulus, however, one may
still linearly expand the band dispersion relations around the new Dirac points,
thereby recovering a cone approximation, but now with elliptical sections
at constant energy, as a result of the strain-induced deformation. This may be
interpreted in terms of robustness of the peculiar quantum state characterizing
the electron liquid in graphene, and can be described as an instance of `quantum
protectorate'. For increasing strain, two inequivalent Dirac points may merge
into one, which usually occurs at either midpoint $M_\ell$ ($\ell=1,2,3$) of the
first Brillouin zone boundary, depending on the strain direction. This
corresponds to the breaking down of linearity of the band dispersions along a
given direction through the Dirac points, the emergence of low-energy
quasiparticles with an anisotropic massive low-energy spectrum, and the opening
of a gap in the energy spectrum. Besides, we confirm that such an event depends
not only on the strain modulus, but characteristically also on the strain
direction. In particular, no gap opens when strain is applied along the armchair
direction.

We derived the energy dependence of the density of states, and recovered a
linear dependence at low energy within the cone approximation, albeit modified
by a renormalized strain-dependent slope. In particular, such a slope
has been shown to increase with increasing strain modulus, regardless of the
strain direction, thus suggesting that applied strain may obtain a steeper DOS
in the linear regime, thereby helping in increasing the carrier concentration of
strained samples, \emph{e.g.} by an applied gate voltage. We have also
calculated the DOS beyond the Dirac cone approximation. As is generic for
two-dimensional systems, the proximity to ETTs gives rise to (possibly
degenerate) Van~Hove singularities in the density of states, appearing as
logarithmic peaks in the DOS.

Finally, we generalized our previous results for the optical conductivity
\cite{Pellegrino:09} to the case of strained graphene. We studied the frequency
dependence of the longitudinal optical conductivity as a function of strain
modulus and direction, as well as of field orientation. Our main results are
that (a) logarithmic peaks appear in the optical conductivity at sufficiently
high frequency, and can be related to the ETTs in the electronic spectrum under
strain, and depending on the strain direction; (b) the relative weight of the
peaks in general depends on the strain direction and field orientation, and
contributes to the generally anisotropic pattern of the optical conductivity as
a function of field orientation; (c) the opening of a band gap, where allowed,
is signalled by a vanishing optical conductivity. Thus, an experimental study of
the optical conductivity in the visible range of frequencies as a function of
strain modulus and direction, as well as of field orientation, should enable one
to identify the occurrence of the three distinct ETTs predicted for graphene \cite{Pellegrino:09c-mod}.

In our study, we have assumed that the chemical potential does not itself depend
on strain. In the doped case, this is clearly an approximation, as the carrier
concentration is expected to remain constant, while the band structure is
modified by strain. It will therefore of interest, for future studies, to
investigate the strain-dependence of the chemical potential required to maintain
a constant carrier concentration. This will enable one to evaluate the
dependence of the Hall resistivity on uniaxial strain, which is a quantity of
experimental interest \cite{Zotos:00,Prelovsek:01}.

\appendix

\begin{figure}[h]
\centering
\includegraphics[height=0.9\columnwidth,angle=-90]{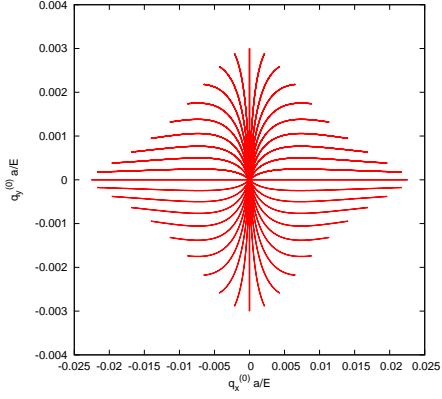}
\caption{(Color online) Showing the scaled position of the center
$(q_{x0}/E,q_{y0}/E)$ of an elliptical section of the Dirac cone around $\bk_D$
at constant energy $E$, Eq.~(\ref{eq:q0}). Each line refers to a given strain
angle $\theta$ with respect to the lattice $x$ axis, and varying strain modulus
$\varepsilon=0-0.2$. \emph{E.g.,} the directions $\theta=0$ and $\theta=\pi/6$
correspond to the vertical bottom and horizontal left line, respectively.}
\label{fig:q0}
\end{figure}

\section{Section of a Dirac cone for strained graphene}
\label{app:ellipse}

Eq.~(\ref{eq:canonical}) yields the canonical form of the ellipse obtained as a
section with constant energy $E$ of the cone approximating the band dispersions
around either Dirac point $\bk_D$, Eq.~(\ref{eq:cone}). The center
$(q_{x0},q_{y0})$ with respect to $\bk_D$ of the ellipse evolves linearly with
energy $E$ according to
\begin{subequations}
\label{eq:q0}
\begin{eqnarray}
q_{x0} &=& \frac{1}{2} A^2 (d_x \cos\eta - d_y \sin\eta) E \\
q_{y0} &=& \frac{1}{2} B^2 (d_x \sin\eta + d_y \cos\eta) E .
\end{eqnarray}
\end{subequations}
The ellipse semiaxes $A$, $B$ are given by
\begin{subequations}
\begin{eqnarray}
\frac{1}{A^2} &=& \frac{1}{2} (\gamma - \sqrt{\alpha^2 + \beta^2} ) \\
\frac{1}{B^2} &=& \frac{1}{2} (\gamma + \sqrt{\alpha^2 + \beta^2} ) .
\end{eqnarray}
\end{subequations}
In the above equations, we have made use of the following definitions:
\begin{subequations}
\begin{eqnarray}
\cos \eta &=& \frac{1}{\sqrt{2}} \left( 1 + \frac{\alpha}{\sqrt{\alpha^2 +
\beta^2}} \right)^{1/2} \\
\sin \eta &=& \frac{\beta}{2\cos \eta \sqrt{\alpha^2 + \beta^2}} \\
\alpha 
       &=& - \frac{3a^2}{2} (t_1^2 + t_2^2 -2t_3^2 ) \\
\beta 
       &=& - \frac{3\sqrt{3}a^2}{2} (t_1^2 - t_2^2 ) \\
\gamma 
       &=& \frac{3a^2}{2} (t_1^2 + t_2^2 + t_3^2 ) ,
\end{eqnarray}
\end{subequations}
where $\bd$ is given in Eq.~(\ref{eq:hK}). One finds $\alpha,\beta\to0$, while
$\gamma\to 9t^2 a^2 /2$ in the limit of no strain, $\varepsilon\to0$.
Fig.~\ref{fig:q0} shows the dependence on the strain modulus of the scaled
coordinates of the ellipse center for different strain orientations.

\begin{small} \bibliographystyle{apsrev}
\bibliography{a,b,c,d,e,f,g,h,i,j,k,l,m,n,o,p,q,r,s,t,u,v,w,x,y,z,zzproceedings,Angilella}
\end{small}

\end{document}